# Bridging the Gap Between Modern UX Design and Particle Accelerator Control Room Interfaces

Rachael Hill[1], Casey Kovesdi[1], Dr. Torrey Mortenson[1], Madelyn Polzin[2], Zachary Spielman[1], and Dr. Katya Le Blanc[1]

[1]Idaho National Laboratory, Idaho Falls, ID 83415, USA
[2]Fermi National Accelerator Laboratory, Batavia, IL 60510, USA

**ABSTRACT**

Accelerator control systems often represent relatively complex and safety-sensitive human–machine interfaces within process control industries. These systems are technically robust and reflect the cumulative integration of solutions built and adapted across decades. One of the regular, unfortunate casualties of provisional accelerator control system updates is their human-system interfaces (HSIs) which often lag behind modern usability and design standards. An additional challenge is that although there is a multitude of established human factors (HF), and user experience (UX) principles for everyday digital applications, there are very few (if any) established principles for complex and safety-critical applications for an accelerator. This paper argues for the importance of established HF and UX principles (herein referred to as *human-centered design principles*) into the development of accelerator HSIs, emphasizing the need for clarity, consistency, responsiveness, and cognitive accessibility. Drawing from HF/UX best practices and human-centered design, this paper discusses how these approaches can enhance operator performance, reduce human error, and improve accelerator personnel collaboration. Case studies from Accelerator Control Operations Research Network (ACORN) at Fermilab are explored to demonstrate how interfaces built with human-centered design principles can scale with system complexity while remaining intuitive and efficient for diverse user roles including operators, machine experts, and engineers. By bridging the gap between traditional control system design and modern human-centered design methods, this paper provides a roadmap for evolving accelerator HSIs into more usable, maintainable, and effective tools.

## INTRODUCTION

Particle accelerators are among the most technically demanding machines ever built, combining vast physical infrastructures with layers of software and hardware control. Their applications span frontier physics experiments, cancer treatment via proton therapy, materials testing, and even national security. Operating these machines requires orchestrating thousands of interdependent subsystems, from vacuum pumps and power supplies to beam diagnostics and radiation safety interlocks.

The human-system interface (HSI) serves as the operator's central lens into this ecosystem. The HSI supports every decision from adjusting beam current, responding to a trip, or coordinating with machine experts flows through the interface. Unlike consumer applications, HSIs in accelerator control rooms are not optional conveniences; they are mission-critical. A poorly designed HSI can obscure essential data or delay a response, potentially jeopardizing expensive equipment or even personnel safety. Conversely, a well-designed HSI enhances situational awareness, supports rapid troubleshooting, and enables collaboration across disciplines. Thus, HSIs are not passive representations of machine status; rather, they enable active engagement in the safe and reliable operation of the accelerator.

Despite their importance, accelerator HSIs have historically evolved in an ad hoc fashion. At Fermilab for instance, the accelerator control network (ACNET) control system dates back to the 1980s and has been incrementally extended ever since. Each new accelerator upgrade, subsystem addition, or experimental need often led to the creation of bespoke panels and displays. Due to the nature of these extensions, solutions were typically optimized for immediate needs rather than long-term usability, which was a reasonable approach given the Laboratory's finite resources and the pressing priority of keeping accelerators running.

As a result, operators often encounter inconsistent layouts, redundant or conflicting data views, and high cognitive load in everyday operations. These issues not only erode efficiency but also amplify risk. Human error is more likely when interfaces are inconsistent, cluttered, or misaligned with user expectations. In practice, experienced operators compensate through extensive training and tacit knowledge, but this reliance on expertise makes onboarding difficult and institutional memory fragile. In other words, even though operators develop exceptional expertise through years of training and experience, the systems often pull their focus away from applying that expertise to operational scenarios and instead force them to wrestle with the intricacies of the interface itself.

In contrast, fields outside of accelerator operations have witnessed a transformation over the past decades. Human factors (HF) and user experience (UX) have matured into established disciplines, underpinned by empirical research, best practices, and widely adopted standards. Human factors is the scientific discipline concerned with understanding the interactions between people and complex systems, with the goal of improving safety, performance, and user well-being. Congruently, UX has emerged as a specialized approach focused on how individuals perceive, navigate, and interact with digital systems with an emphasis on usability. Together, HF and UX offer a rich toolkit for creating interfaces that are not only functional but also intuitive, adaptive, and most importantly, user-centric.

The Accelerator Control Operations Research Network (ACORN) project at Fermilab highlighted how HF/UX methods and a structured HSI style guide can modernize accelerator control interfaces to improve usability, safety, and long-term maintainability (Hill et al., 2024). With the establishment of ACORN, there is now an opportunity to approach HSI development more strategically. By integrating human-centered design from the outset, upgrades can be guided not only by technical necessity but also by long-term usability, scalability, and operator effectiveness.

## Overview of Accelerator Control Network (ACNET)

ACNET is Fermilab's long-standing control system, originally developed in the early 1980s to coordinate the laboratory's growing suite of accelerators. Built on a distributed, client–server architecture, ACNET integrates thousands of devices including power supplies, vacuum systems, beam instrumentation, and safety interlocks into a unified operational environment. Its design emphasizes robustness and extensibility, allowing it to evolve alongside successive accelerator upgrades over more than four decades. Through a combination of command-line tools, graphical panels, and data logging services, ACNET provides operators and other accelerator personnel with real-time monitoring, control, and diagnostic capabilities. However, while ACNET has proven remarkably durable and technically reliable, its interfaces reflect the incremental and patchwork development characteristic of legacy systems where HSIs often vary in layout, visual style, and interaction patterns, and new features have typically been added in response to immediate operational demands rather than guided by long-term usability goals. As such, ACNET stands as both a testament to engineering resilience and a case study in the HF and UX challenges of sustaining complex control systems over decades of operation.

## Overview of Human Factors, UX, and Usability in Design

As previously discussed, HF is the scientific study of human characteristics, abilities, and limitations in relation to interacting with work systems or products with a focus in ensuring that the systems or products are usable by human operators. Often the phrase "fitting the work to the worker" is used in human factors engineering (HFE). User Experience (UX) is a discipline focused on the design and evaluation of digital systems, products, and services. Usability, often considered a component within UX, emphasizes functional effectiveness and task performance; however, UX encompasses a broader spectrum of considerations, including aesthetic perception, and overall satisfaction with an interaction. In this sense, UX integrates both the objective requirements of usability and the subjective dimensions of experience, positioning itself as the primary discipline for analyzing and shaping how people perceive, interact with, and ultimately value designed systems. These domains focus on a 'human-centered' approach across the discipline which seeks to place real people at the center of the development process. By employing human-centered approaches, we can ensure that the eventual system or product that is built has considered the needs of the people who will make use of it.

There are many different principles in the broader digital design realm but there is some foundational guidance to ensuring a successful design, in general. Principles such as Nielsen's Ten Heuristics (Nielsen, 1994), Cooper's Interaction Design paradigms (Cooper, 2014), Gestalt Design Principles (Wertheimer, 1923; Koffka, 1935; Ellis, 1938; and Chang, et al., 2002 for a digital application), Norman's foundational design text *The Design of Everyday Things* (Norman, 2013), and Shneiderman's 8 Golden Rules of Interface Design (Shneiderman, 2005). These are just a handful of the most cited and foundational frameworks used in human-centered design. Additionally, there is application specific guidance as well, for example in nuclear power NUREGs -0711 and -0700 (O'Hara, et al., 2012; O'Hara and Fleger, 2020) and aviation (Ahlstrom, 2016).

All these principles and guidelines are valuable resources for the design of different systems and for different user populations. A robust understanding and command of these concepts can have significant impact on a designer's ability to design a functional and satisfying user experience.

## BENEFITS OF USER EXPERIENCE AND USABILITY DESIGN

Integrating HF and UX principles can drive critical benefits to the development process at accelerator facilities, particularly human-centered approaches. There are many different benefits and opportunities, however this section will focus on four primary areas: improved performance and situation awareness; reduction of human error likelihood; faster onboarding and decreased training burden; and, increased flexibility and maintainability of the developed systems.

These are four areas that capture specific needs across accelerator facilities that HF and UX can provide. A final benefit to the early and often integration of these disciplines is the decreased development cost of these systems and increased efficiencies or revenues generated as a result. Research and case studies from consultancy McKinsey and Co. captured critical aspects of the value of design (Sheppard, et al., 2018). By integrating these efforts early, development teams can minimize future rework, capture more clear user requirements, and develop systems that are likely to be adopted.

### Improved Performance and Situation Awareness

Applying HF and UX principles enables interfaces that highlight critical information, reduce clutter, and provide intuitive visual hierarchies. Clear visual cues improve real-time situation awareness, helping operators prioritize attention in dynamic environments (Burns, et al., 2008). These benefits are significant in safety critical industries such as accelerators where the industrial and economic risk can be significant. By developing interfaces that achieve high levels of human performance and a robust awareness of the system, organizations can create more resilient operational frameworks and achieve efficiencies.

### Reduction of Human Error

Consistency in layout and interaction patterns reduces cognitive overhead, allowing operators to focus on decision-making rather than interface navigation. Error-preventive design—confirmation prompts, visual warnings, constrained input—further safeguards many industrial operations (Nielsen, 1994). One of the most significant impacts of human-centered design is the reduction of human error . By designing the interfaces to 'fit the worker' rather than trying to force the user into using a poorly designed interface we can increase the likelihood of task success and minimize error traps and create operator support features for more complex tasking (Norman, 2013). The reduction of human error should be a high-level focus for operations organizations at accelerator facilities due to the significant cost of the equipment, processes, and potential research achievements (Sheppard, et al., 2018). Many of the facilities are bespoke and therefore represent a higher economic risk to organizations. By adopting a human-centered approach that risk can be mitigated.

### Faster Onboarding and Lower Training Burden

Legacy HSIs often require months of training. Interfaces informed by HF and UX principles can lower onboarding time by leveraging familiar design conventions, reducing reliance on institutional knowledge and easing transitions for new staff (Burke and Hutchins, 2007). Accelerator facilities have extremely complex and varied experimental needs which demands a highly trained and highly adaptable workforce for successful operation. Significant efficiencies can be gained by leveraging robust design principles in the development of operational interfaces (Norman, 2013; Cooper, 2014). Better designs will more closely match user expectations as to functionality and performance and will be developed with users in mind. A core aspect of these design efforts is the natural intuitiveness of the designed interface. A system which embodies this intuitiveness from a foundational level will require less training effort and time for operators to become proficient and allow them to focus on building the resiliency and adaptability for the accelerator context (Cooper).

### Increased Flexibility and Maintainability

User requirements that are derived from robust user research efforts can be incredibly impactful for development teams that are often juggling many different requirements or demands from stakeholders or systems. User requirements are the most fundamental requirement for a digital system and clear user requirements give development teams a coherent and effective path forward in their efforts. Further, system technical requirements can be integrated into the user research process as well which can help developers have their needs heard as they build the interface. This close integration is key for long term maintenance and support for a digital control system. Additionally, a modular, design-system-driven HSI approach allows incremental evolution. Interfaces can be adapted for new roles, machines, or experiments without complete overhauls. This scalability supports long-term sustainability of control systems. The first step in this process (i.e., the development of an accelerator context specific HSI style guide) was initially completed in 2023 (Hill et al., 2023). For more information on this, see Recommendation 2.

## RECOMMENDATIONS FOR INTERGRATION

The ACORN at Fermilab has begun experimenting with HF/UX-driven redesigns of accelerator HSIs. Our recommendations include the following principles:

- Involve End Users Early
- Establish a Style Guide and Design Philosophy

- Develop a Design Library
- Enable Continuous and Systematic Involvement

These principles are based on a collection of experiences in applying HF/UX methods (herein referred to as human-centered methods) to the ACORN project over the course of three years. We discuss each principle in detail next.

**Recommendation 1: Involve End Users Early**

Championing the end users by involving them as early as possible into the project lifecycle is a core tenet to human-centered design (e.g., ISO 9241-112:2025). A key motivation for this is to ensure that the design of the HSIs within the control system are *needs based*, whereby their requirements are directly informed by the end user's needs in safely, effectively, and efficiently completing their tasks. Their needs were ultimately identified in the project using human-centered methods like interviews, observations, and task analysis; they were then carefully considered and synthesized with human-centered design principles that informed key project outputs such as the HSI style guide and design library.

A second motivation for getting end users involved early was to enable 'stakeholder buy in' to the modernization process and new control system. Oftentimes end users like operators may be hesitant to change because they are extensively familiar with the design of their existing HSIs and may not necessarily see the value of a large-scale change in the design of the HSIs. To address this, the HF/UX design team included end users like operators at the onset of the project where they ultimately gained experience in aspects of the new design through structured human-centered design activities to become *champions* in the design of the HSIs. With getting end users involved early and identifying existing pain points, we have observed that this strategy can ultimately break the resistance to change while giving them opportunity to be proactive in designing HSIs that fit for them.

**Recommendation 2: Establish a Human-System Interface Style Guide**

The HSI style guide provides a clear, consistent framework for the design and development of the HSIs. The style guide uses guiding principles and concepts, as previously highlighted, to offer design guidance for the HSIs associated with the accelerator control system. In our case, it was treated as a 'living document' in the sense that revisions were issued at different points in the project. These revisions reflected updates to its guidance based on expanding the scope of different roles using the HSIs. For instance, a first instance of the style guide was developed with guidance focused primarily on the operators of the accelerator control system, despite the project having identified other roles (e.g., main control room crew chiefs, accelerator physicists, various engineers and technicians, etc.). This was intentional to help prioritize and focus development of HSIs for the operators while the project concurrently performed early human-centered contextual research for these other roles.

The style guide explicitly discussed its scope and highlighted where gaps remained. By the next iteration of the style guide, updates to the document were then made after completing contextual research for the roles, among making updates based on other design decisions made across the project, such as choosing a user interface development toolkit. In any case, important takeaways to emphasize from this work is that:

1) A HSI style guide should be developed for the project and managed by the HF/UX design team.
2) It should be treated as a living document in which updates to it are possible throughout the lifecycle of the accelerator control system.
3) The guidance that is provided in the style guide is grounded in three key elements: user needs, established design principles and guidance, as well as the native characteristics of the selected hardware and software.

**Recommendation 3: Develop a Design Library**

A design library (oftentimes referred to as an *HSI toolkit*) is a collection of the graphical symbols and other elements needed to implement the HSI style guide (ANSI/ISA-101.01-2015). In this project, the HSI style guide was a key input into developing a design library where the design guidance offered in the style guide are thereby reflected as 'implementation ready' instances of the guidance. Examples of elements in the design library consist of specific accelerator soft controllers, graphs, tables, labels, or navigation buttons to name a few.

A key advantage of developing a design library, based on our experiences in its application through ACORN, is that it provides a common design framework that can be readily used at implementation. The design library ensured consistency across instances of different display elements used throughout the control system HSIs while also reducing unnecessary complexity in maintaining each element. By developing a single common instance of a particular display element, modifications to that element can be done only once in which all instances of that element become modified. This offers notable benefits in maintainability and efficiency throughout the project lifecycle.

**Recommendation 4: Enable Continuous and Systematic Involvement**

Finally, an important element in successfully integrating human-centered design (i.e., the application of HF and UX principles) into the accelerator modernization project lifecycle entails continuous and systematic involvement throughout the modernization project lifecycle. Such involvement of keeping end users embedded in the modernization process through focused activities that elicit their expertise, experiences, and overall feedback ultimately reduces assumptions made when developing design artifacts such as the style guide and design library.

The HF/UX design team initiated systematic user research activities by completing a comprehensive task analysis for the operators of the Fermilab accelerator. This activity spanned a course of multiple months where 15 operators provided operational insights with their roles, responsibilities, and tasks performed. This was accomplished using a mixed-methods approach, combined with semi-structured interviews and virtual screenshare capabilities to observe tasks demonstrated through structured methods such as walkthrough analysis. The results of this contextual research provided technical bases into the initial revision to the HSI style guide, in combination with use of known HF and UX standards and guidelines (e.g., ISO 11064-5:2008, ISO 9241-112:2025, and ANSI/HFES-100:2007), as well as use of original research results (e.g., Cooper et al., 2014; Endsley, 2004, etc.).

Next, the HF/UX design team systematically expanded the scope of the HSI design to additional roles across the Fermilab accelerator complex. Over the course of a year, the team performed a series of semi-structured interviews with personnel of various technical roles such as physicists, engineers, and technicians, as well as supervisory roles such as crew chiefs, division directors, and managers. Each interview collected contextual information of each role including their level of experience and education, their primary jobs and tasks performed, level of training, communication and coordination requirements, and key software applications used to perform their tasks.

The information was synthesized and used to further inform key project artifacts including the style guide and design library. Moreover, a detailed application analysis was performed to establish a mapping between every accelerator application used (pointing to the ones impacted by the modernization), key roles using the application, and overarching function that the application supports for the accelerator. An output of this activity offered a graph visualization of the interdependencies between applications to fulfill specific accelerator functions while also mapping the specific roles responsible. This information provided an effective approach to 1) ensuring that all applications were being accounted for in the modernization process as well as 2) identifying potential opportunities for consolidating applications in the new control system to reduce unnecessary complexity. This work further has scope to perform iterative usability tests with prototypes that reflect the design guidance provided from the style guide and design library.

**CONCLUSION**

Human–system interfaces constitute a critical yet comparatively underdeveloped element of accelerator control systems. Although legacy designs demonstrate significant technical robustness, their usability limitations can constrain operator effectiveness, and hinder collaboration across multi-disciplinary users. This paper has advanced the case for systematically integrating HF and UX design principles into accelerator HSIs, drawing upon case studies from Fermilab's ACORN initiative to illustrate both feasibility and benefit.

Embedding HF and UX methodologies within the culture of accelerator operations enables the development of interfaces that are not only technically sound but also scalable, intuitive, and cognitively accessible, thereby aligning more closely with the needs of diverse accelerator users, most notably operators. To realize this vision, institutional commitment is essential: investment in HF and UX integration must be recognized as a foundational design priority, positioning HSIs not as provisional or ancillary components but as central enablers of safe, efficient, and sustainable accelerator operations.

## ACKNOWLEDGMENT

This manuscript has been authored by Battelle Energy Alliance, LLC under Contract No. DE-AC07-05ID14517 with the U.S. Department of Energy. The United States Government retains and the publisher, by accepting the article for publication, acknowledges that the U.S. Government retains a nonexclusive, paid-up, irrevocable, world-wide license to publish or reproduce the published form of this manuscript, or allow others to do so, for U.S. Government purposes.